\documentclass[prl,aps,preprintnumbers,twocolumn]{revtex4}

\usepackage{epsfig}
\usepackage{latexsym}

\newcommand{\be}{\begin{equation}} 
\newcommand{\ee}{\end{equation}} 
\newcommand{\bea}{\begin{eqnarray}} 
\newcommand{\eea}{\end{eqnarray}} 
\newcommand{\mpl}{M_{\rm P}}

\begin{document}

\preprint{SU-GP-03/6-2}

\title{Is Cosmic Speed-Up Due to New Gravitational Physics?}

\author{Sean M. Carroll$^1$\footnote{carroll@theory.uchicago.edu}, 
Vikram Duvvuri$^1$\footnote{duvvuri@theory.uchicago.edu}, 
Mark Trodden$^2$\footnote{trodden@physics.syr.edu} 
and Michael S. Turner$^{1,3,4}$\footnote{mturner@oddjob.uchicago.edu}} 
\affiliation{$^1$Enrico Fermi Institute, Department of Physics, 
and Center for Cosmological Physics, University of Chicago, 
5640 S. Ellis Avenue, Chicago, IL 60637-1433, USA. \\ 
$^2$Department of Physics, Syracuse University, 
Syracuse, NY 13244-1130, USA. \\ 
$^3$Department of Astronomy \& Astrophysics, University of Chicago, 
Chicago, IL 60637-1433, USA. \\ 
$^4$NASA/Fermilab Astrophysics Center, 
Fermi National Accelerator Laboratory, Batavia, IL 60510-0500, USA.}

\begin{abstract} 
We show that cosmic acceleration can arise due to very tiny
corrections to the usual gravitational action of general relativity of
the form $R^n$, with $n<0$. This eliminates the need for dark energy,
though it does not address the cosmological constant problem. Since a
modification to the Einstein-Hilbert action of the form $R^n$, with
$n>0$, can lead to early-time inflation, our proposal provides a
unified and purely gravitational origin for the early and late time
accelerating phases of the Universe. 
\end{abstract}


\maketitle

{\em Introduction.} That the expansion of the Universe is currently
undergoing a period of acceleration now seems inescapable: it is
directly measured from the light-curves of several hundred type Ia
supernovae~\cite{Riess:1998cb,Perlmutter:1998np,Tonry:2003zg}, and
independently inferred from observations of the cosmic microwave
background (CMB) by the WMAP satellite~\cite{Bennett:2003bz} and other
CMB experiments~\cite{Netterfield:2001yq,Halverson:2001yy}.

Cosmic speed-up can be accommodated within general relativity by
invoking a mysterious cosmic fluid with large negative pressure,
dubbed {\em dark energy}.  The simplest possibility for dark energy is
a cosmological constant; 
unfortunately, the smallest estimates for its value
are 55 orders of magnitude too large (for reviews see
\cite{Carroll:2000fy,Peebles:2002gy}).  This fact has motivated a host
of other possibilities, most of which assume a zero cosmological
constant, with the dynamical dark energy being associated
with a new scalar
field~\cite{Wetterich:fm,Ratra:1987rm,Caldwell:1997ii,Armendariz-Picon:1999rj,
Armendariz-Picon:2000dh,Armendariz-Picon:2000ah,Mersini:2001su,Caldwell:1999ew,
Carroll:2003st,Sahni:1999gb,Parker:1999td}.

However, none of these suggestions is compelling and most have serious
drawbacks.  Given the challenge of this problem, it is worthwhile
considering the possibility that cosmic acceleration is not due to
some kind of stuff, but rather arises from new gravitational
physics~\cite{Deffayet:2001pu,
Freese:2002sq,Ahmed:2002mj,Arkani-Hamed:2002fu,Dvali:2003rk}. This is
the angle we pursue in this letter.

General relativity is based upon the Einstein-Hilbert action, with
Lagrange density $\sqrt{-g}R$, where $R$ is the curvature scalar.  A
natural modification is to add terms to the
action that are proportional to $\sqrt{-g}R^n$.  It is known that, for
$n>1$, such terms lead to modifications of the standard cosmology at
early times which lead to de Sitter behavior (Starobinskii
inflation~\cite{Starobinsky:te}).  Here, we show that, for $n<0$, such
corrections become important in the late Universe and can lead to
self-accelerating vacuum solutions, providing a purely gravitational
alternative to dark energy.

The self-accelerating solutions we find can behave like vacuum energy
(i.e., $w_{\rm DE}\equiv P_{\rm DE}/\rho_{\rm DE} = -1$), or can lead
to power-law acceleration with cosmic scale factor $a \propto t^q$,
$q>1$, and $w_{\rm DE} < -2/3$. We also argue that our model is
consistent with existing tests of gravitation theory and that the
accelerating phase is consistent with current cosmological
observations.

{\em A Model.}  Many authors have considered modifying the
Einstein-Hilbert action with terms that become effective in the
high-curvature region. Here, however, we explore modifications which
become important at extremely low curvatures to explain cosmic
speed-up. For definiteness and simplicity we focus on the simplest
correction to the Einstein-Hilbert action, 
\begin{equation}
\label{action} S =\frac{\mpl^2}{2}\int d^4 x\,
\sqrt{-g}\left(R-\frac{\mu^4}{R}\right) +\int d^4 x\, \sqrt{-g}\,
{\cal L}_M \ . 
\end{equation} 
Here $\mu$ is a new parameter with units
of $[{\rm mass}]$, ${\cal L}_M$ is the Lagrangian density for matter
and the reduced Planck mass $M_P\equiv (8\pi G)^{-1/2}$.  
For work on theories where the Lagrangian takes the form
$f(R)$, see \cite{Barrow:rx,Barrow:xh,Barrow:hg,Magnano:bd,dobado,
Schmidt:gb,Capozziello:2003tk}.

The field equation for the metric is then 
\begin{eqnarray} \label{fieldequation}
 \left(1+\frac{\mu^4}{R^2}\right)R_{\mu\nu} -
 \frac{1}{2}\left(1-\frac{\mu^4}{R^2}\right)Rg_{\mu\nu}\qquad \qquad  &&
 \nonumber \\ 
 + \mu^4\left[g_{\mu\nu} \nabla_{\alpha}\nabla^{\alpha}
 -\nabla_{(\mu}\nabla_{\nu)}\right]R^{-2}
  =\frac{T_{\mu\nu}^M}{\mpl^2} \ ,&&
\end{eqnarray} 
where $T_{\mu\nu}^M$ is the matter energy-momentum
tensor.

The constant-curvature vacuum solutions, for which $\nabla_{\mu}R=0$,
satisfy $R=\pm\sqrt{3}\mu^2$. Thus, we find the interesting result
that {\it the constant-curvature vacuum solutions are not Minkowski
space, but rather are de Sitter space and anti de Sitter space}. This
bifurcation of the vacuum solutions from a unique one (Minkowski
space) to two new ones (de Sitter and Anti-de Sitter spaces) as
$\mu^4$ is increased from zero is unusual and is to be contrasted with
the case of a cosmological constant. We will see that the de Sitter
solution is, in fact, unstable, albeit with a very long decay time
$\tau\sim \mu^{-1}$.

We consider a perfect-fluid energy-momentum tensor,
\begin{equation} \label{perfectfluid}
T_{\mu\nu}^M = (\rho_M + P_M)U_{\mu} U_{\nu} + P_M g_{\mu\nu}\ ,
\end{equation} 
where $U^{a}$ is the fluid rest-frame four-velocity,
$\rho_M$ is the energy density, $P_M$ is the pressure and we write
$P_M=w\rho_M$. Matter corresponds to $w=0$ and radiation to $w=1/3$.
We take the metric to be of the 
flat Robertson-Walker form, $ds^2=-dt^2+a^2(t)d{\bf x}^2$,
for which the curvature scalar satisfies
\begin{eqnarray} R &=&
6\left[\frac{\ddot a}{a}+\left(\frac{\dot a}{a}\right)^2\right] =
6\left(\dot H + 2 H^2 \right) \ ,
\end{eqnarray}
where an overdot denotes differentiation with respect to time,
$H=\dot a/a$ and $a(t)$ is the scale factor.  

The time-time component of the field equations for this metric is
\begin{eqnarray} \label{newfriedmann} 3H^2 &-& \frac{\mu^4}{12({\dot
H}+2H^2)^3}\left(2H{\ddot H} \right. \nonumber \\ &+& \left.15H^2{\dot
H}+2{\dot H}^2+6H^4\right) = \frac{\rho_M}{\mpl^2}\ .
\end{eqnarray}
This replaces the usual Friedmann equation, recovered by setting $\mu=0$.
The space-space components of (\ref{fieldequation}) lead to the other
independent Einstein equation,
\begin{eqnarray} \label{newspacespace} &&{\dot H} +
\frac{3}{2}H^2 - \frac{\mu^4}{72({\dot H}+2H^2)^2} \left[4{\dot
H}+9H^2 \right. \nonumber \\ 
&-& \left. R^2\partial_0 \partial_0
\left(\frac{1}{R^2}\right) -2R^2 H\partial_0
\left(\frac{1}{R^2}\right)\right] =-\frac{P_M}{2\mpl^2} \ . 
\end{eqnarray}

The above fourth-order equations are complicated and it is
difficult to extract details about cosmological evolution from them.
It is convenient to transform from our current frame, which we call
the {\em matter frame}, to
the {\em Einstein frame}, where the gravitational Lagrangian takes
the Einstein-Hilbert form and the
additional degrees of freedom ($\ddot H$
and $\dot H$) are represented by a fictitious scalar field $\phi$.

Following~\cite{Magnano:bd}, we make a conformal transformation
\begin{equation} \label{conformaltrans} 
{\tilde g}_{\mu\nu}=p(\phi)g_{\mu\nu} \ , \ \ \ p\equiv \exp
\left(\sqrt{\frac{2}{3}}\frac{\phi}{\mpl} \right) \equiv
1+\frac{\mu^4}{R^2} \ , 
\end{equation} 
where $\phi$ is a real scalar
function on space-time, $d{\tilde t}=\sqrt{p}dt$, and ${\tilde
a}(t)=\sqrt{p} a(t)$. It is then convenient to define an
Einstein-frame matter energy-momentum tensor by 
\begin{equation}
\label{confperfectfluid} {\tilde T}_{\mu\nu}^M = ({\tilde\rho}_M +
{\tilde P}_M){\tilde U}_{\mu} {\tilde U}_{\nu} + {\tilde P}_M {\tilde
g}_{\mu\nu}\ , \end{equation} where ${\tilde U}_a\equiv \sqrt{p}U_a$,
${\tilde\rho}_M=\rho_M/p^2$ and ${\tilde P}_M = P_M/p^2$.

In terms of the new metric ${\tilde g}_{\mu\nu}$, our theory is that
of a scalar field $\phi(x^{\mu})$ minimally coupled to Einstein
gravity, and non-minimally coupled to matter, with potential
\begin{equation} \label{potential} V(\phi)=\mu^2 \mpl^2
\frac{\sqrt{p-1}}{p^2} \ , 
\end{equation} 
shown in the figure
below. Note that $V(\phi )\rightarrow 0$ both for $\phi\rightarrow 0$
($p\rightarrow 1$) and for $\phi \rightarrow \infty$ ($p\rightarrow
\infty$) and achieves its maximum, $9\mu^2\mpl^2/16 \sqrt{3}$, at $p =
4/3$. 
\begin{figure}[htb] \label{isotropy} \centerline{
\psfig{figure=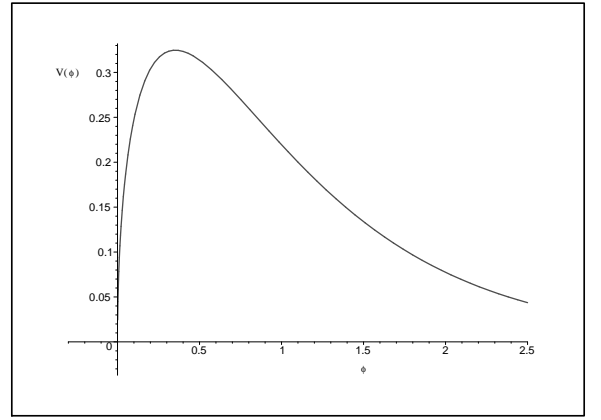,height=3in,angle=270}} \caption{The
Einstein-frame potential $V(\phi)$} 
\end{figure}

Denoting with a tilde all quantities (except $\phi$) in the Einstein frame, 
the relevant Einstein-frame cosmological equations of motion are
\begin{equation}
 \label{conffriedmann}
 3{\tilde H}^2 = \frac{1}{\mpl^2}\left[\rho_\phi 
 + {\tilde\rho} \right] \ ,\\ \label{confscalar}
\end{equation}
\begin{equation}
 \phi''+3\tilde H\phi'+\frac{dV}{d\phi}(\phi)- \frac{(1-3w)}{\sqrt{6}}
 {\tilde\rho}_M =0 \ , 
\end{equation} 
where a prime denotes $d/d{\tilde t}$, and where 
\begin{equation}
 {\tilde\rho}_M = \frac{C}{{\tilde a}^{3(1+w)}}
 \exp\left[-\frac{(1-3w)}{\sqrt{6}}\frac{\phi}{\mpl} \right] \ , 
\end{equation} 
with $C$ a constant, and 
\begin{equation}
 \rho_\phi = \frac{1}{2}\phi'^2  + V(\phi) \ . 
\end{equation} 
The matter-frame Hubble parameter $H$ is related to that in the Einstein 
frame ${\tilde H}\equiv {\tilde a}'/{\tilde a}$ by 
\begin{equation}
 H = \sqrt{p} \left({\tilde H} -{\phi' \over \mpl\sqrt{6}}\right) \ . 
\end{equation}

What can we say about cosmological solutions in the more friendly
setting of the Einstein frame? Ordinarily, Einstein gravity with a
scalar field with a minimum at $V=0$ would yield a Minkowski vacuum
state. However, here this is no longer true. Even though $V\rightarrow
0$ as $\phi\rightarrow 0$, this corresponds to a curvature singularity
and so is not a Minkowski vacuum. The other minimum of the 
potential, at $\phi\rightarrow \infty$, is clearly not a static
solution. These statements are also easily
seen in the matter frame.

Now let us first focus on vacuum cosmological solutions, i.e.,
$P_M=\rho_M = 0$. The beginning of the Universe corresponds to $R
\rightarrow \infty$ and $\phi \rightarrow 0$.  The initial conditions
we must specify are the initial values of $\phi$ and $\phi'$, denoted
as $\phi_i$ and ${\phi'}_i$.  For simplicity we take $\phi_i \ll
\mpl$. There are then three qualitatively distinct outcomes, depending
on the value of ${\phi'}_i$.

{\em 1.  Eternal de Sitter.}  There is a critical value of ${\phi'}_i
\equiv {\phi'}_C$ for which $\phi$ just reaches the maximum of the
potential $V(\phi)$ and comes to rest.  In this case the Universe
asymptotically evolves to a de~Sitter solution (ignoring spatial
perturbations).  As we
have discovered before (and is obvious in the Einstein frame), this
solution requires tuning and is unstable, since any perturbation will
induce the field to roll away from the maximum of its potential.

{\em 2.  Power-Law Acceleration.}  For ${\phi'}_i > {\phi'}_C$, the
field overshoots the maximum of $V(\phi )$. Soon thereafter, the
potential is well-approximated by $V(\phi) \simeq \mu^2\mpl^2
\exp(-\sqrt{3/2}\phi /\mpl)$, and we easily solve for ${\tilde
a}({\tilde t}) \propto {\tilde t}^{4/3}$, which corresponds to $a(t)
\propto t^2$ in the matter frame. Thus, the Universe evolves to
late-time power-law inflation, with observational consequences similar
to dark energy with equation-of-state parameter $w_{\rm DE}=-2/3$.

{\em 3.  Future Singularity.}  For ${\phi'}_i < {\phi'}_C$, $\phi$
does not reach the maximum of its potential and rolls back down to
$\phi =0$.  This yields a future curvature singularity in which the
Hubble parameter remains finite in both the matter and Einstein
frames, but its time derivative becomes infinite because $\dot H
\propto V^\prime \rightarrow 1/\sqrt{\phi} \rightarrow \infty$.

We now turn to the more interesting case in which the Universe
contains matter. As can be seen from~(\ref{confscalar}), the major
difference here is that the equation-of-motion for $\phi$ in the
Einstein frame has a new term.  Furthermore, since the matter density
is much greater than $V \sim \mu^2\mpl^2$ for $t \ll 14\,$Gyr, this
term is very large and greatly affects the evolution of $\phi$.  The
exception is when the matter content is radiation alone ($w=1/3$), in
which case it decouples from the $\phi$ equation due to
conformal invariance. 

Despite this complication, it is possible to show that the three
possible cosmic futures identified in the vacuum case remain in the
presence of matter. To see this, first note that, if we ignore the
$V'$ term in~(\ref{confscalar}) because it is subdominant at early
times, and assume that $3{\tilde H}^2 = {\tilde \rho}_M/\mpl^2$, we
obtain 
\begin{equation} {\phi'}({\tilde t}) = {\phi'}({\tilde
t}_0)\left(\frac{{\tilde a}_0}{{\tilde a}}\right)^3 +
\frac{C}{\sqrt{6}}\left(\frac{{\tilde t}-{\tilde t}_0}{{\tilde
a}^3}\right) \ , \end{equation} for $w=0$, and \begin{equation}
{\phi'}({\tilde t}) = {\phi'}({\tilde t}_0)\left(\frac{{\tilde
a}_0}{{\tilde a}}\right)^3 \ , 
\end{equation} 
for $w=1/3$.

The value of ${\phi'}$ redshifts as it would for a free scalar field
and, for $w=0$, increases dramatically due to the presence of matter.
By tuning $\phi'_i$, the value of $\phi^\prime$ at the epoch when
matter becomes unimportant (i.e., ${\tilde \rho}_M < \mu^2\mpl^2$) can
be adjusted, resulting in the same three futures as in the vacuum
theory.

Thus far, we have left the dimensionful parameter $\mu$ unspecified
beyond the requirement that it be sufficiently small that precision
tests of gravity theory are not affected (see below).
By choosing $\mu\sim 10^{-33}\,$eV, the corrections to the standard
cosmology only become important at the present epoch, making our
theory a candidate to explain the observed acceleration of the
Universe without recourse to dark energy. It is, of course, important
to be clear that we have merely {\it chosen} the value of $\mu$ to
achieve this end.  Since we have no particular reason for this choice,
such a tuning appears no more attractive than the traditional choice
of the cosmological constant.  However, it is intriguing to note that,
in the present circumstance, the smallness of the $1/R$ term in the
action is directly related to the lateness of the accelerating
phase. Thus, even an extremely tiny correction to the Einstein action
can eventually have dramatic consequences.

Clearly our choice of correction to the gravitational action can be
generalized.  Terms of the form $-\mu^{2(n+1)}/R^n$, with $n>1$, lead
to similar late-time self acceleration. Such actions can be
transformed into the Einstein frame with scalar field potential
\begin{equation} \label{generalpot} V(p) = \mu^2\mpl^2
\left(\frac{n+1}{2n}\right) n^{1/(n+1)} \frac{(p-1)^{n/(n+1)}}{p^2} \ , 
\end{equation} 
with resulting matter-frame scale factor $a(t)\propto
t^q$, where 
\begin{equation} \label{genscale} q =
\frac{(2n+1)(n+1)}{n+2} \ . 
\end{equation} 
Such a modification thus
yields behavior similar to a dark energy component with equation of
state parameter 
\begin{equation} \label{geneos} w_{\rm eff} = -1 +
\frac{2(n+2)}{3(2n+1)(n+1)} \ . 
\end{equation} 
As $n\rightarrow\infty$
the expansion approaches an exponential and the space-time is
approximately de Sitter. Clearly therefore, such modifications can
easily accommodate current observational
bounds~\cite{Melchiorri:2002ux,Spergel:2003cb} on the equation of
state parameter $-1.45< w_{\rm DE} <-0.74$ ($95\%$ confidence level).

Finally, any modification of the Einstein-Hilbert action must, of
course, be consistent with the classic solar system tests of gravity
theory, as well as numerous other astrophysical dynamical tests. Many
such tests depend on the Schwarzschild solution, which Birkhoff's
theorem ensures is the unique, static, spherically-symmetric
solution. It is simple to prove a related result here: that the most
general, static, spherically-symmetric solutions are the
Schwarzschild-(anti) de Sitter solutions.
These solutions differ, of course, from the familiar Schwarzschild
solution, which asymptotically approaches Minkowski space. However, it
is clear that, if the parameter $\mu^4$ is chosen small enough, any
astrophysical tests of gravity that depend on the Schwarzschild
solution will be unaffected by the modification we have made. We
expect similar arguments will hold for the cases of charged black
holes and those with angular momentum.

{\em Concluding Remarks.}  We have shown that the current epoch of
cosmic acceleration can arise through purely gravitational effects,
eliminating the need for dark energy. In particular, modifications to
the Einstein--Hilbert action of the form $-\mu^{2(n+1)}/R^n$, with
$\mu \sim 10^{-33}\,$eV and $n<0$, can lead to cosmic speed up with
$-1< w_{\rm eff} \leq -2/3$.

There is, of course, growing evidence from measurements of CMB
anisotropy, as well as of large-scale structure, that the Universe
also evolved through a period of accelerated expansion very early on,
known as inflation.  It is also known that modifications to the
Einstein-Hilbert action of the form $m^{-2}R^2$, with $m$ a mass
parameter, yield an Einstein-frame potential 
\begin{equation} \label{} V (\phi ) = m^2\mpl^2\frac{(p - 1)^2}{8p} \ , 
\end{equation} 
where,
as before, $p\equiv \exp [\sqrt{2/3}\phi /\mpl ]$. In this case, the
beginning of the Universe occurs at large $\phi$ and the Universe
undergoes inflation. Thus, our work implies that it is possible for
both eras of cosmic acceleration to arise from the gravitational
sector of the theory, with general relativity being a valid
description of the Universe only at intermediate cosmic times.

There are certainly many other details of this theory that remain to
be checked. In particular, we would like to analyze quantitatively the
behavior of the transition from matter domination to self acceleration
as well as the growth of linearized perturbations about the given
background solutions we have identified.

\acknowledgments We thank C. Armendariz-Picon, T. Beloreshka,
M. Bowick and Laura Mersini for extensive discussions and
R. Brandenberger, E. Lim, D. Marolf, R. Myers, T. Jacobson and
T. Vachaspati for helpful comments. The work of SMC is supported in
part by the DOE, the NSF, and the Packard Foundation.  VD is supported
in part by the NSF.  MT is supported in part by the NSF under grant
PHY-0094122, and is a Cottrell Scholar of Research Corporation.  MST
is supported in part by the US DOE (at Chicago), the NASA (at
Fermilab), and the NSF (at Chicago).

\end{document}